\documentclass{article}
 
\usepackage{glas}

\usepackage{wrapfig}
 
\begin{document}

\begin{titlepage}{GLAS-PPE/2010-06}{28$^{\underline{\rm{th}}}$ June 
2010}

\title{Prompt photon production at HERA}

\author{D.H. Saxon$^1$\\
$^1$ University of Glasgow, Glasgow, G12 8QQ, Scotland\\
\\
{\em On Behalf of the H1 and ZEUS collaborations}}

\begin{abstract}
 
New results are presented on prompt photon production in 
photoproduction (H1) and DIS (ZEUS). These are compared to the predictions 
of collinear (DGLAP) factorisation and $k_T$-factorisation theories. The 
comparison tends to favour the $k_T$-factorisation approach.

\vspace{1.5 cm}
\begin{center}
{\em XVIII International Workshop on Deep-Inelastic 
Scattering and
 Related Subjects\\}
{\em  April 19 -23, 2010\\
 Convitto della Calza, Firenze, Italy}
\end{center}
\end{abstract}
\vspace{2 cm}
\begin{center}
 Thanks to Tobias Haas for presenting this at the 
conference\\
 while I was marooned in Scotland by the volcano 
Eyjafjallajoekull.
\end{center}
\newpage
\end{titlepage} 
 
 \section{Isolated photons}
 New results are presented on prompt photon production in photoproduction 
\cite{H1photo} and deep inelastic scattering \cite{ZDIS}. The data arise 
from 320 to $340~{\rm pb}^{-1}$ of HERA-II collisions, including both 
$e^+p$ and 
$e^-p$ data sets  These are used 
to test the predictions of various theories with collinear factorisation 
or $k_T$-factorisation. The comparison at low $Q^2$ and low $x$ tends to 
favour the $k_T$-factorisation approach.

High-$E_T$ isolated photon emission offers a new and reliable 
probe of dynamics in $e^{\pm}$-proton collisions. The photon is the only 
stable final state 
particle that couples to the quark line in the Feynman diagram. For this 
reason theorists refer to it as a `prompt' photon (not coming from hadron 
decay). Experimentally one observes `isolated' signals in the detector.
In DIS events there are two hard scales, the 
$Q^2$ of the exchanged photon and the $E_T$ of the observed photon.  
The observed photon can be radiated from a quark line (refered to as the 
QQ process) or the lepton line (LL process). The interference term (LQ 
process) is small for isolated photons and also changes sign between 
$e^+p$ and $e^-p$ collisions. The LQ term is therefore neglected in this 
work and $e^+p$ and $e^-p$ data sets are combined. Note that in the LL 
process the electron recoils against the high-$E_T$ photon into the 
detector acceptance and the event is 
therefore classified as DIS. As a result the 
prompt-photon photoproduction data set contains only the QQ process. In 
this case the Feynman diagram can involve direct exchanged photons 
(coupling via the process $\gamma q \rightarrow \gamma q$) or resolved 
exchanged photons, which couple to gluons in the proton via the process
 $g q \rightarrow \gamma q$.)  

Two stages are involved in extracting the isolated-photon signal. First 
the final-state electron is removed and the remaining calorimeter 
energy-flow objects (EFOs) are clustered into jets, using the 
inclusive-$k_T$ algorithm with parameter $R_0 = 1.0$ \cite{kTjet}. One 
then 
looks at 
the ratio
$R_\gamma = E(\gamma)/E(\gamma -{\rm jet})$ where
$E(\gamma)$ is the energy of the electromagnetic-calorimeter EFO, and $E(\gamma-{\rm jet})$ is the energy of the jet 
that includes the electromagnetic cluster. Demanding $R_\gamma > 0.9$ 
eliminates 
a large fragmentation-dominated background. (The signal lies predominantly 
at $R_\gamma > 0.98$.)

\begin{figure}
\begin{center}
\includegraphics[width=.8\textwidth]{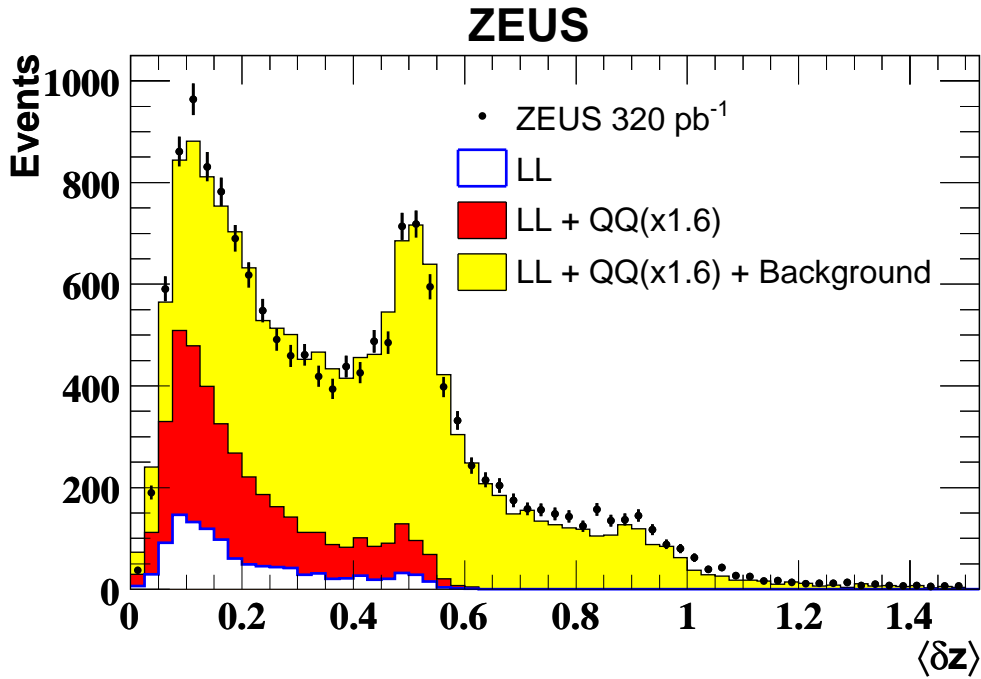}
\caption{ZEUS shower $z$-width distribution}
\end{center}
\label{fig1}
\end{figure}

The isolated electromagnetic EFO thus identified is dominated by 
unresolved clusters of two or more photons, mainly from $\pi^0 \rightarrow 
\gamma \gamma$ decay, which has a minimum opening angle given by $\sin 
\theta/2 = m(\pi^0)/E(\pi^0)$, giving a typical separation of a few cm in 
the electromagnetic calorimeter. One therefore needs to extract a 
narrow-EFO signal for isolated single photons. H1 use a discriminant 
method based on six shower-shape variables - transverse radius, symmetry 
and kurtosis, first layer energy fraction, hot core fraction and hottest 
cell fraction. The ZEUS work reported here uses the fine-granularity 
projective geometry of the barrel electromagnetic calorimeter to 
distinguish $1\gamma$ and $2\gamma$ peaks. Figure \ref{fig1} shows the 
distribution 
in $ \langle \delta Z \rangle = \Sigma E_i | Z_i - 
Z_{cluster}|/(w_{cell} \Sigma E_i)$. The fit is discussed below.

\begin{wrapfigure}{r}{84mm}
\vspace{-2.5cm}
\begin{center}
\includegraphics[width=.56\textwidth]{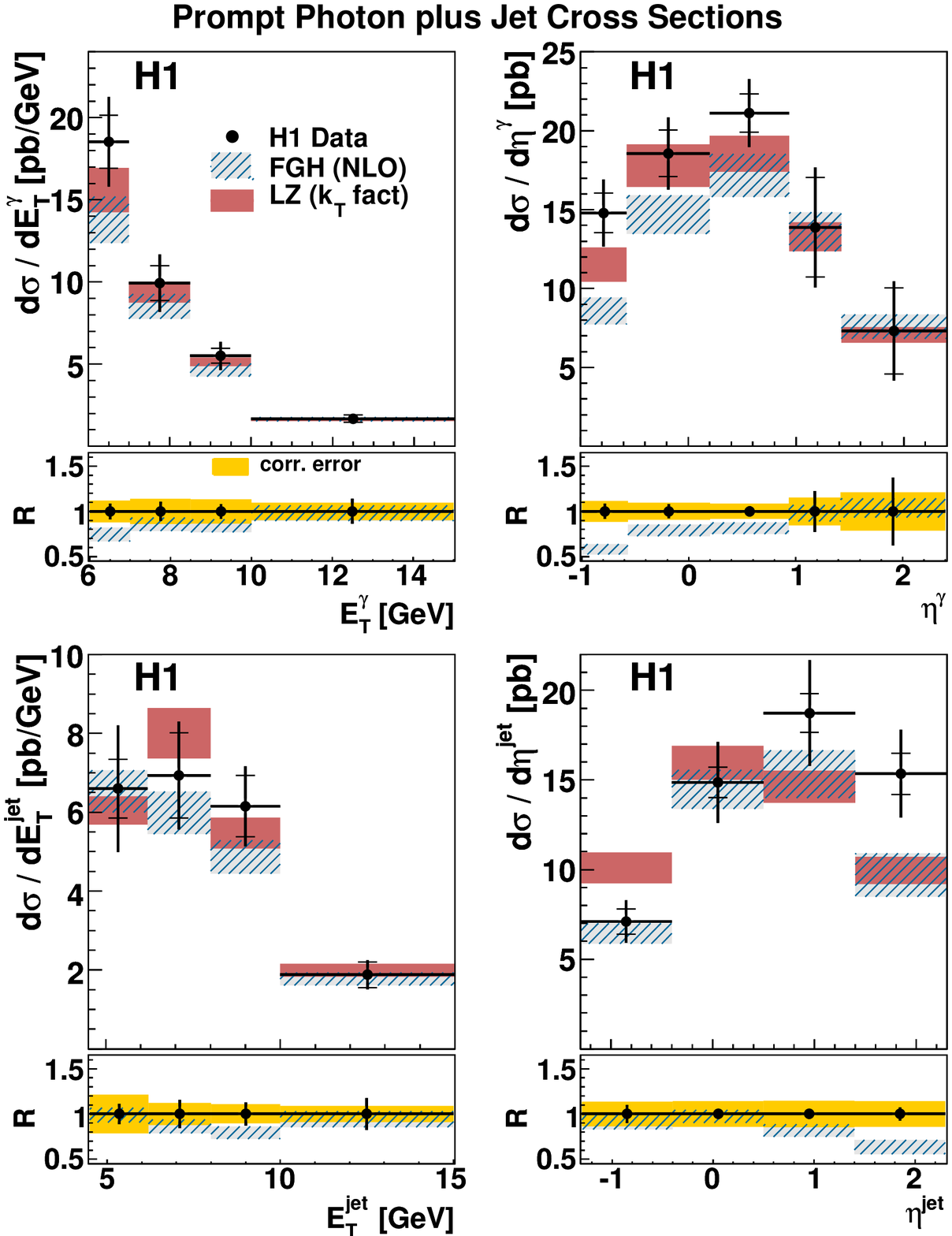}
\includegraphics[width=.56\textwidth]{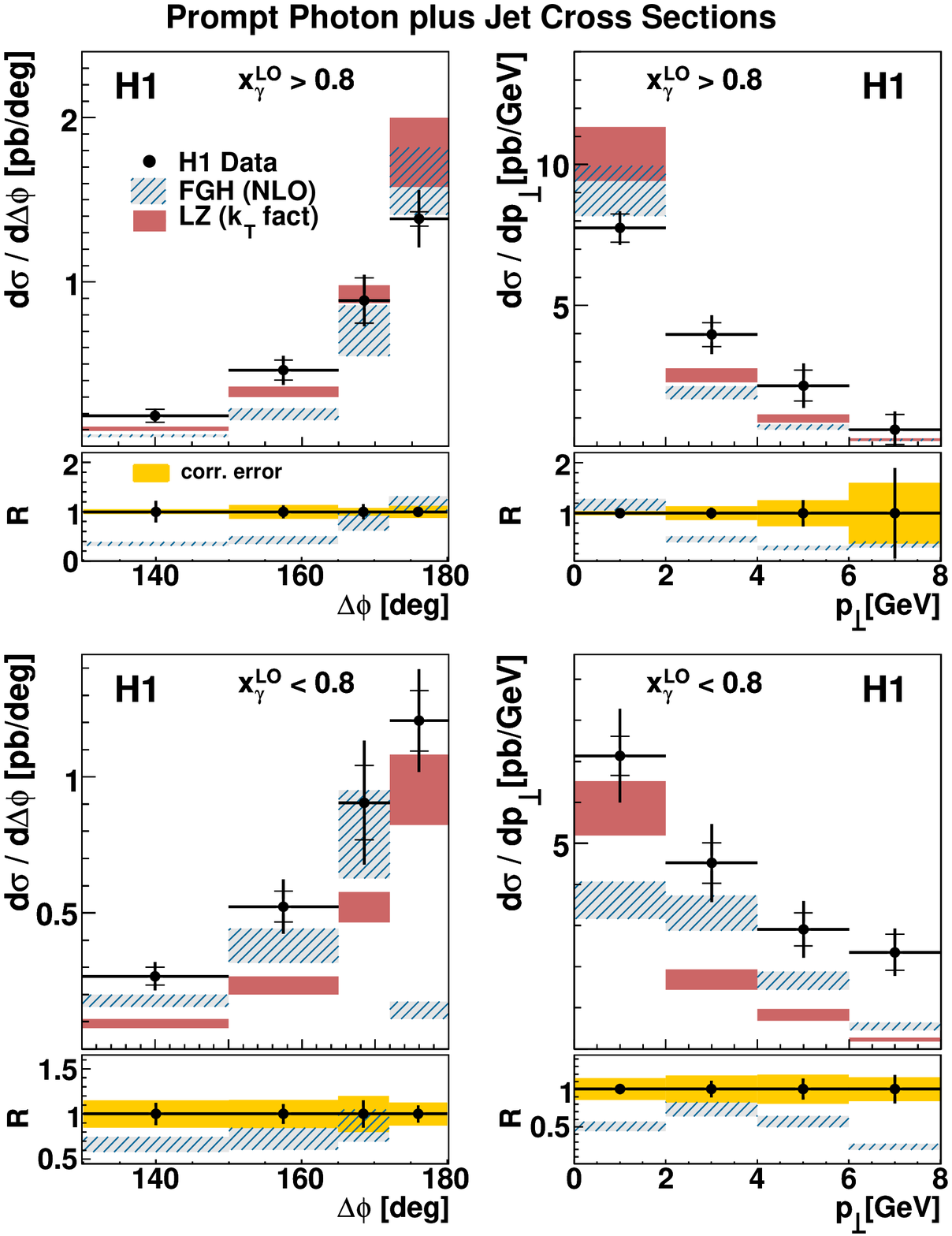}
\caption{H1 results compared to theoretical predictions}
\end{center}
\vspace{-2.7 cm}
\label{fig2}
\end{wrapfigure}

\section{Photoproduction}
H1 present results on inclusive $\gamma$ and $(\gamma+{\rm jet})$ in the 
kinematic range as follows. For the photon $6 < E^T < 15~{\rm GeV}$ and 
$-1.0 < \eta < 2.4$, $R_\gamma > 0.9$, $Q^2 < 1~{\rm GeV}^2$, $0.1< 
y_{JB} < 0.7$ and for jets $E^T({\rm jet}) > 4.5~{\rm GeV}$ and $-1.3 < 
\eta ({\rm jet}) < 2.3$.

They compare their results to NLO QCD. Fontannaz, Guillet and 
Heinrich (FGH) use 
collinear factorisation and DGLAP evolution, include ($O(\alpha_s^2)$) and 
some higher terms and use
MRST01 and AFG2 pdf's for $p$ and $\gamma$ \cite{FGH}. H1 also compare 
their results to a 
$k_T$-factorisation calculation due to Lipatov and Zotov (LZ) \cite{LZ} 
which 
uses direct and resolved integrated parton densities. In comparing the
$(\gamma+{\rm jet})$ data to theory hadronisation corrections (estimated 
using {\sc Pythia}) are of order $8\%$.

Figure 2 includes the transverse energy and pseudorapidity 
distributions for 
the $(\gamma + {\rm jet})$ final state compared to predictions. 
There is some tendency for the photon rapidity to favour LZ. This is also 
seen in earlier ZEUS results \cite{Zphoto}.

H1 separate the photon-plus-jet data into direct and resolved exchanged 
photons using photon and jet directions and the photon energy, (high 
and low values of $x^{LO}_\gamma$), and use 
these 
data sets to study azimuthal photon-jet correlations. In figure 2 $\Delta 
\phi$ is the difference in their azimuth angles and
 $p_\perp$ is the momentum in 
the transverse plane that, if added to the photon, would make its azimuth 
angle opposite to the jet. None of the theories 
describes these data well. Additional 
higher-order ~theory terms could improve the agreement.

\begin{figure}
\centering
\includegraphics[width=.48\textwidth]{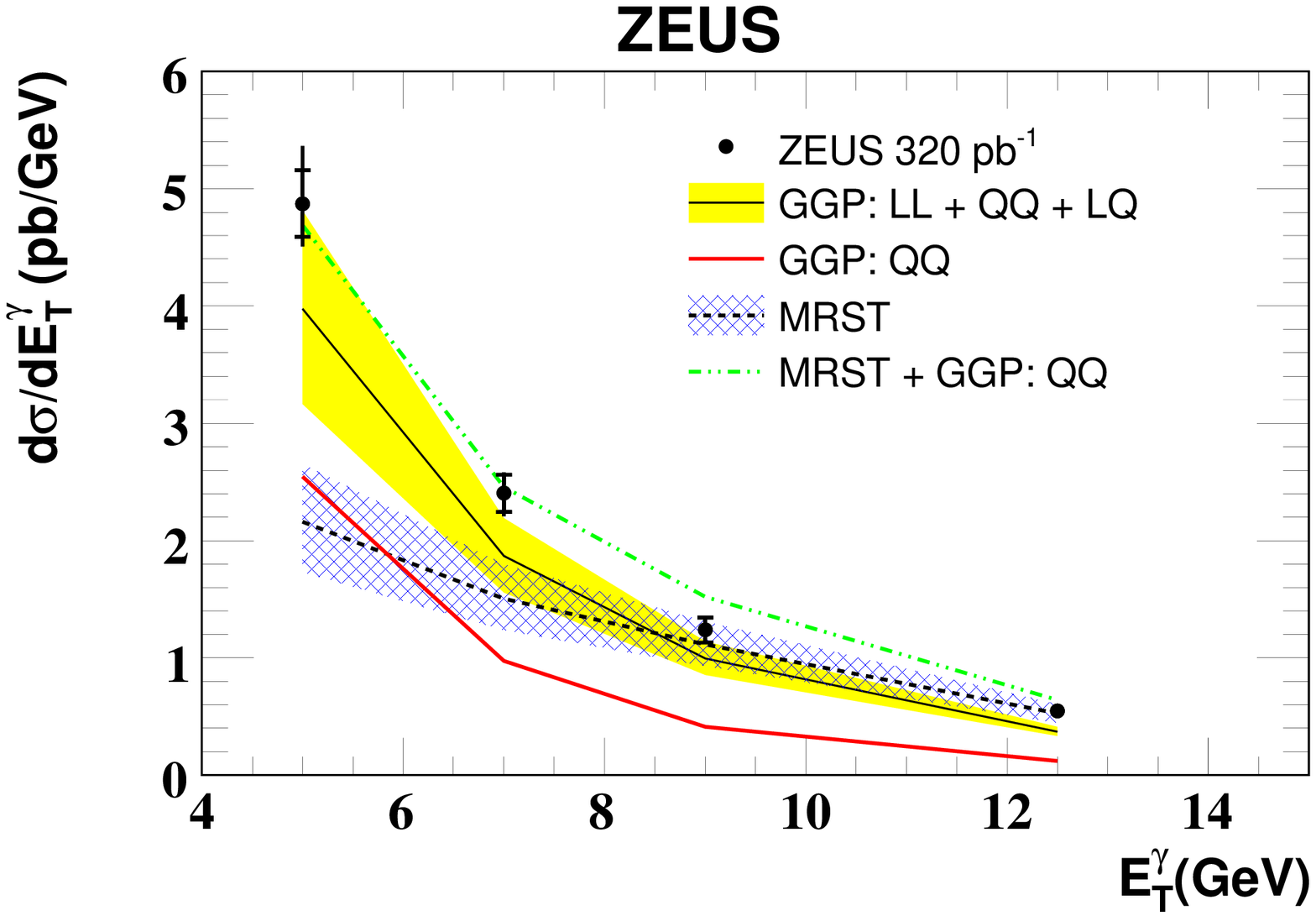}
\includegraphics[width=.48\textwidth]{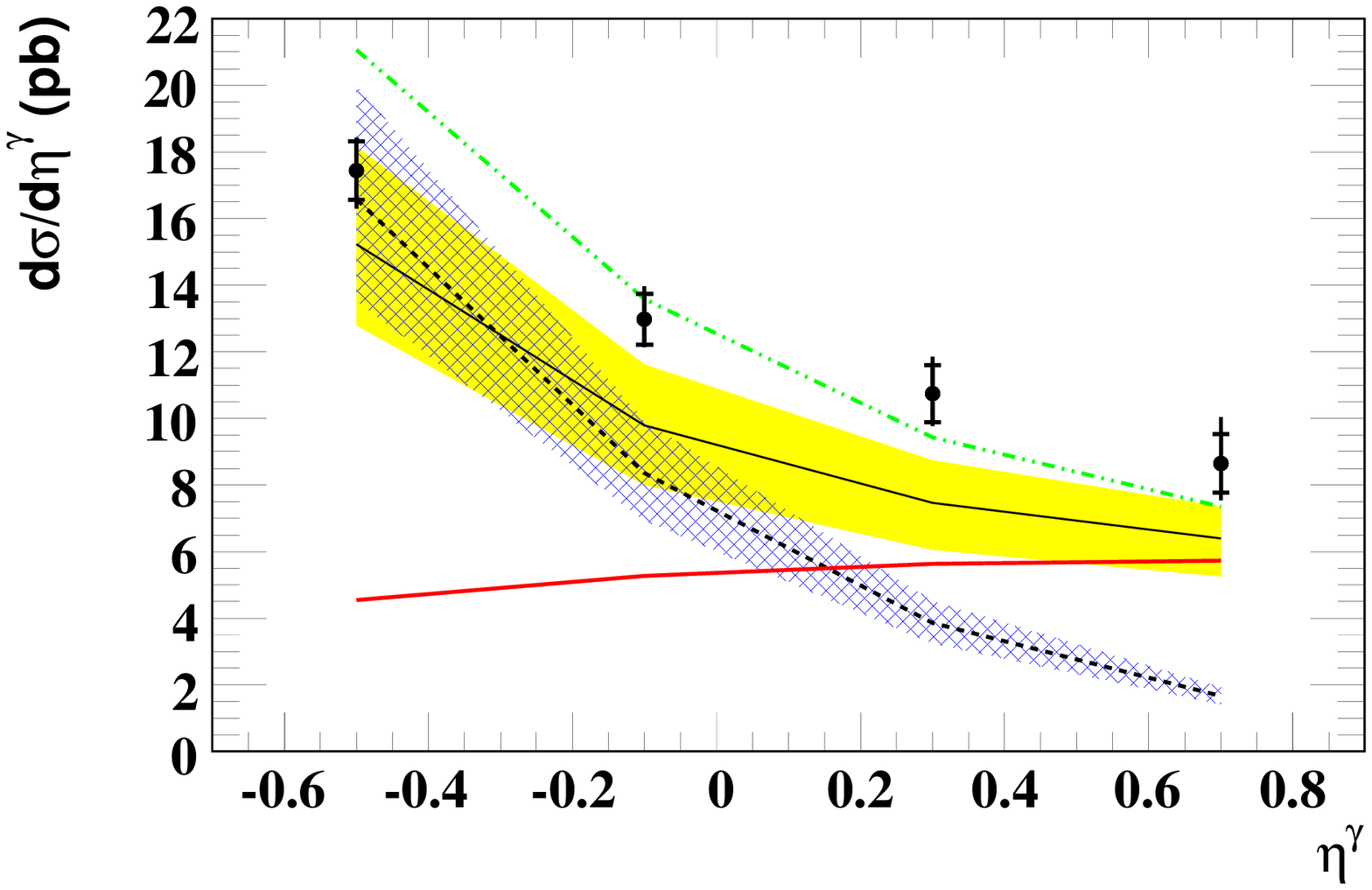}
\includegraphics[width=.48\textwidth]{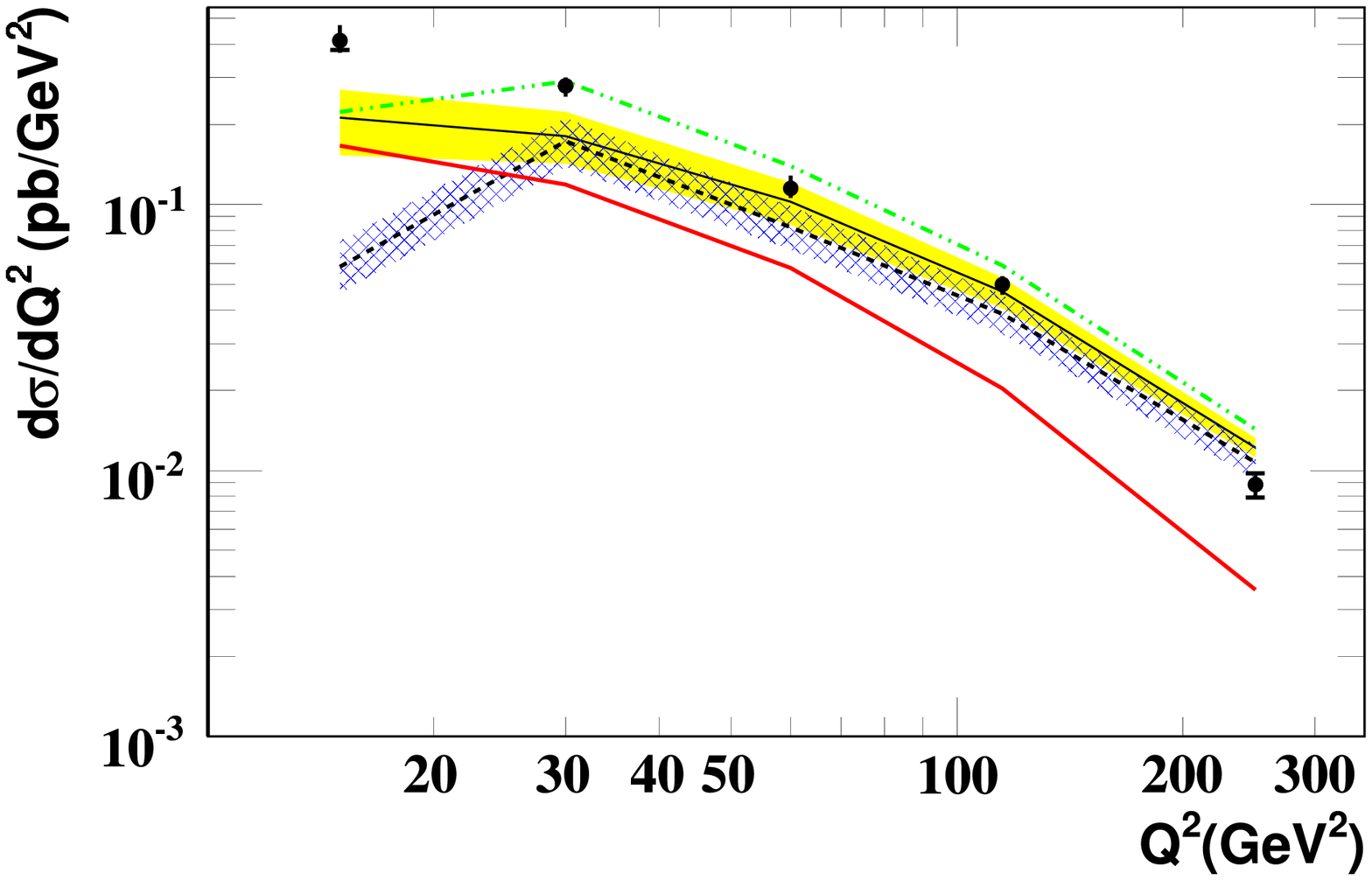}
\includegraphics[width=.48\textwidth]{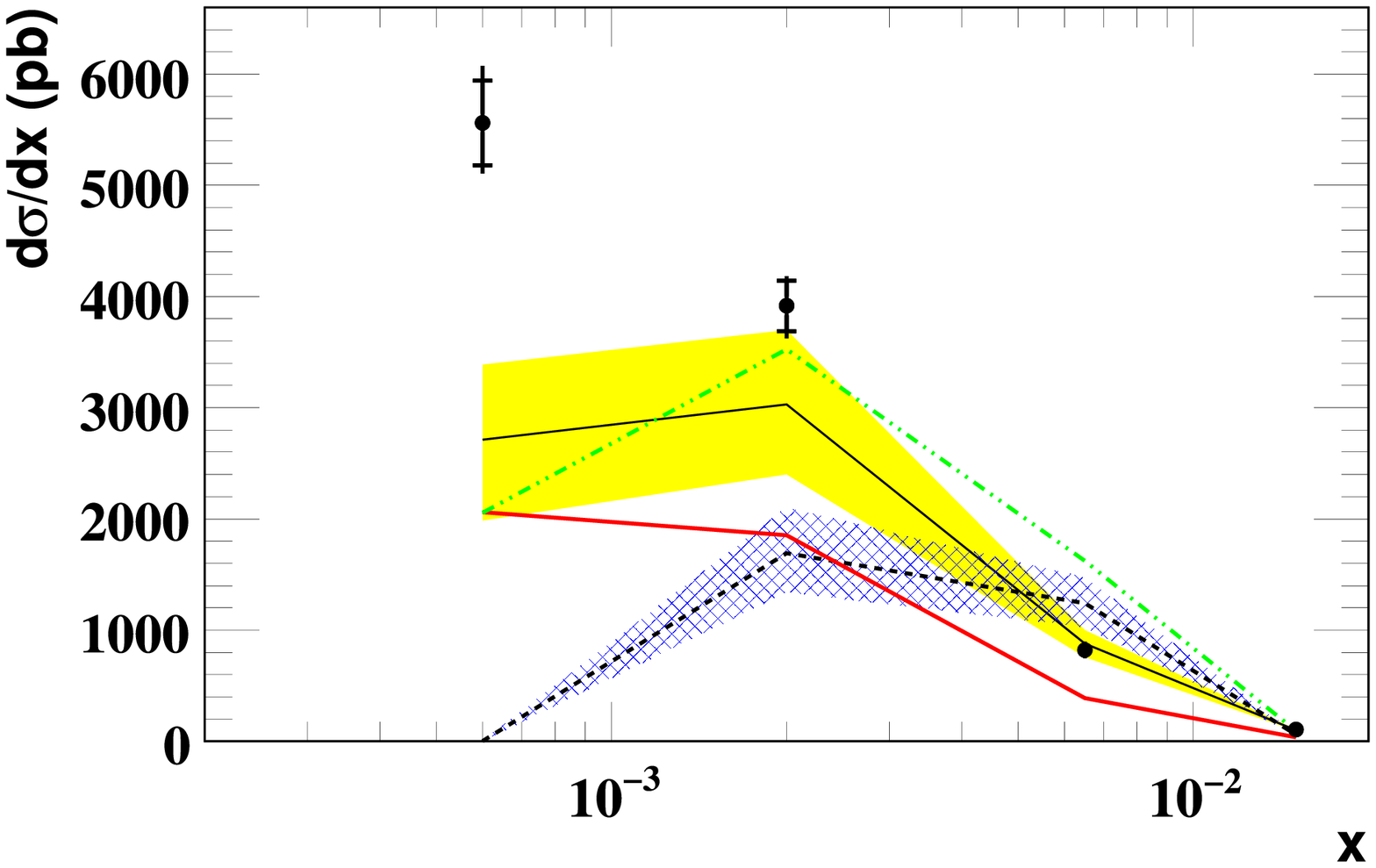}
\caption {ZEUS DIS results and theoretical predictions}
\label{fig4}
\end{figure}

\section{Inclusive prompt photons in DIS}

Figure \ref{fig1} shows the ZEUS isolated photon candidates fitted as the 
sum of LL 
and QQ events, and hadronic background predicted by Monte 
Carlo. In this fit the purely leptonic events are assumed to be accurately 
calculated and the normalisation is held fixed, as is the hadronic 
background. One then needs to rescale the Monte Carlo QQ prediction by 1.6 
to obtain 
the required number of events. This factor was then held fixed in the 
bin-by-bin signal extraction. (For practical reasons The Monte Carlo's 
used were {\sc Pythia}6.416 for QQ and {\sc Django}6-{\sc 
Heracles}4.8.6-{\sc Ariadne} for LL and the hadronic 
background \cite{ZDIS}.) 
Figure \ref{fig4} shows the data 
distributions for photon $E_T$, $\eta$, and for $Q^2$ and $x$.
The photon kinematic range is $4 < E^T < 15 ~{\rm GeV}$ and 
$-0.7 < \eta < 0.9$, $R_\gamma > 0.9$, and $10 < Q^2 < 350~ {\rm GeV}^2$
(measured using the outgoing electron, which must have energy above 
10~GeV 
and lie in the angular range $139.8^\circ$ to $171.8^\circ$.) A cut on the 
final state hadronic mass to be above 5~GeV excludes DVCS events. The 
Monte Carlo 
predictions (not shown) describe $E_T$ and $\eta$ well, but fall below the 
data at low $x$ and at low $Q^2$.

The results in figure \ref{fig4} are compared to a number of theoretical 
predictions.
GGP is the state-of-the-art $O(\alpha^3 \alpha_s)$ collinear factorisation
prediction \cite{GGP}. The production of photons by jet fragmentation is 
included but is suppressed by the isolation cuts. MRST calculate 
$e-\gamma$ collisions 
where the $\gamma$ is part of the photon structure \cite{MRST}. This can 
be 
thought of as the LL process with radiative corrections to all orders. 
We therefore show also a prediction for GGP:QQ+MRST. This gives a 
reasonable description of the data except at low $Q^2$ and at low $x$. 
Earlier H1 
results on $E_T, \eta, Q^2$ show the same behaviour \cite{H1DIS}.

\begin{figure}
\centering
\includegraphics[width=.45\textwidth]{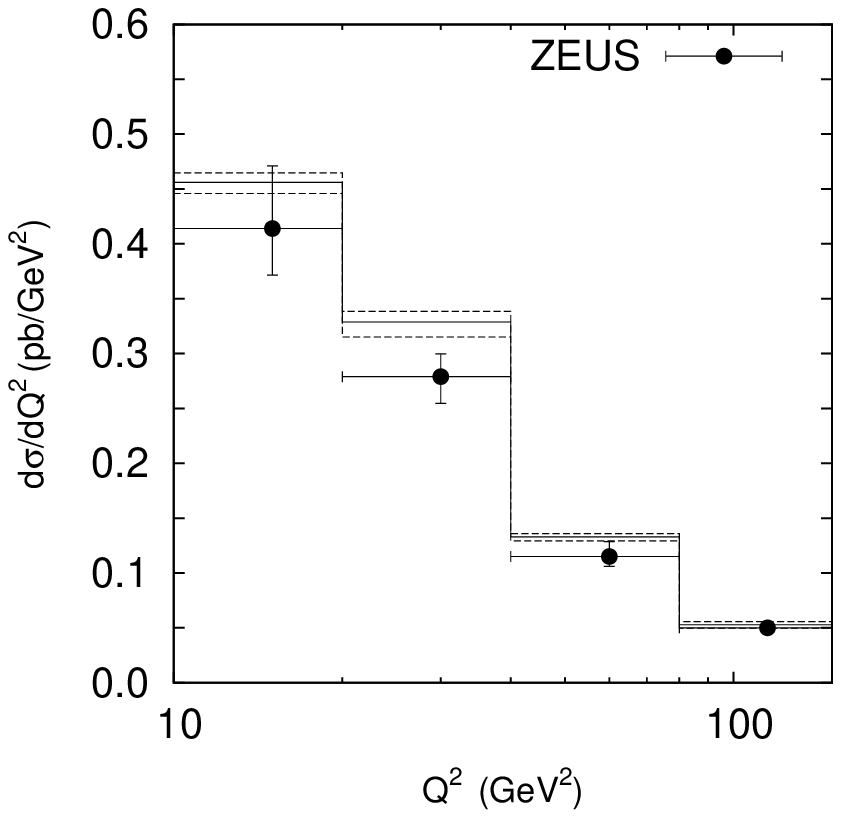}
\includegraphics[width=.45\textwidth]{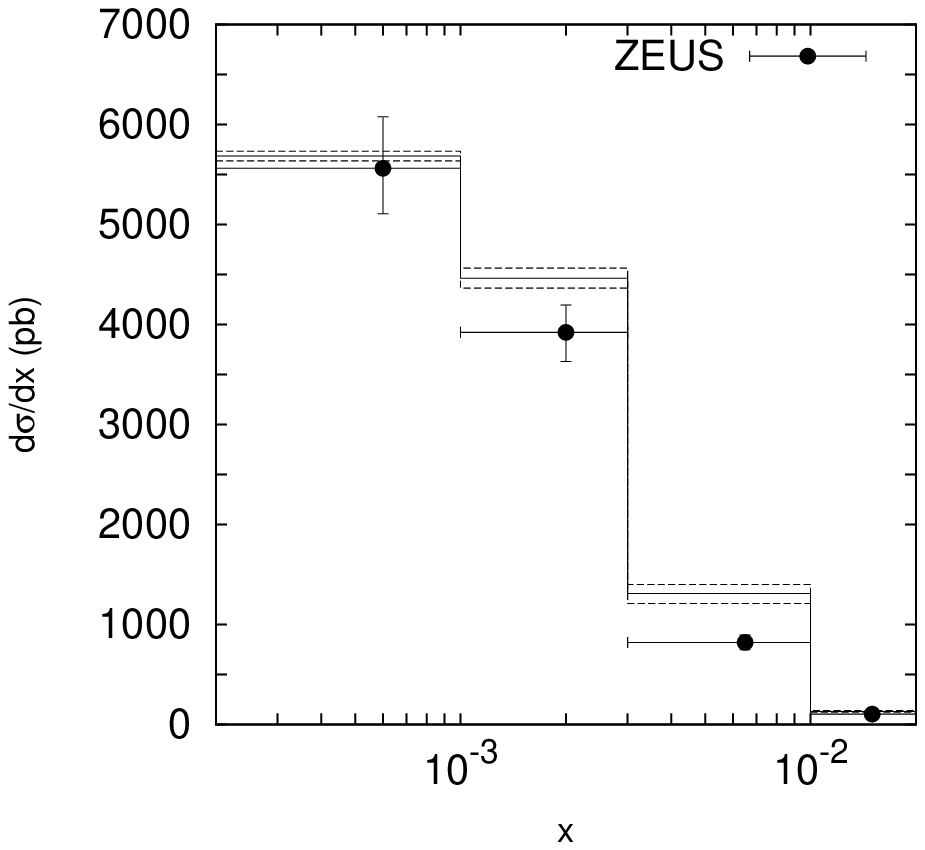}
\caption{Baranov, Lipatov and Zotov predictions for $Q^2$ and $x$ compared 
to ZEUS DIS results}
\label{fig5}
\end{figure}

Baranov, Lipatov and Zotov have calculated inclusive prompt photon 
production using the $k_T$-factorisation approach ($eq* \rightarrow 
eq\gamma$) \cite{BLZ}. Compared to collinear factorisation, this is 
expected the increase the cross-section at low $Q^2$ and low $x$.
The agreement with the data in figure \ref{fig5} is impressive compared 
to figure 3.

Prompt photon production in DIS therefore tends to favour the 
validity of the $k_T$-factorisation approach. Photoproduction results do 
not disagree with this conclusion.

 \end{document}